\pdfoutput=1

\documentclass[10pt,letterpaper]{article}
\usepackage{opex3}
\usepackage{epstopdf}

\begin{document}

\title{Saturation Spectroscopy of Iodine in Hollow-core Optical Fibre}

\author{Anna Lurie$^{1,*}$, Philip S. Light$^1$, James Anstie$^1$,  Thomas M. Stace$^2$, Paul C. Abbott$^1$, Fetah Benabid$^3$, and Andre N. Luiten$^{1}$}

\address{$^1$School of Physics, The University of Western Australia, Australia\\
$^2$Department of Physics, University of Queensland, Brisbane, Queensland 4072, Australia\\
$^3$Centre for Photonics and Photonic Materials, Department of Physics, University of Bath, United Kingdom
\\
}

\email{Andre.Luiten@uwa.edu.au} 



\begin{abstract*}
We present high-resolution spectroscopy of  ${I}_{2}$ vapour that   is  loaded and trapped within the core of a hollow-core photonic crystal fibre (HC-PCF). We compare the observed spectroscopic features to those seen in a conventional iodine cell and show that the saturation characteristics differ significantly.  Despite the confined geometry it was still possible to obtain sub-Doppler features with a spectral width of $\sim 6$\,MHz with very high contrast. We provide a  simple theory which closely reproduces all   the key observations of the experiment. 
 \end{abstract*}

\bibliography{ref}


\section{Introduction}


Iodine vapour has played a prominent role in the frequency stabilization of lasers  over many years because it exhibits a large number of intrinsically narrow spectral lines across a good fraction of the visible and near-IR spectrum.  Some of the best performing vapour-cell frequency standards have used  iodine in large-diameter ($\sim 10$\,cm) and long ($\sim 1$ metre) vapour cells ~\cite{Hall1, Hall2, Argence,40cmcell, 60cmcell, zang}.  The large diameter  minimizes broadening associated with transit of the molecules through the probing laser beam, while the long length allows  operation at low pressures, thereby avoiding collisional broadening, while still demonstrating  strong absorption features.  In order to avoid the substantial challenges   associated with using large and fragile vapour cells there has recently been a great deal of  research targeted towards loading vapours into the cores of specially tailored Hollow-Core Photonic Crystal Fibre (HC-PCF).  Rubidium, ammonia and acetylene  have all been loaded in the core of these fibres~ \cite{FetahScience, FetahNature, gaeta, Ghosh, Hald, PhilRB, Hendrickson, Knabe, chrispaper}, although, to our knowledge, no-one  has yet loaded   iodine in such a fibre and presented detailed spectroscopic measurements.

The key benefit of this  fibre technology is  its robust and compact geometry, which lends itself to easy temperature control and magnetic-shielding, while still possessing the advantage of a long interaction zone (potentially even  longer than the macroscopic glass cells). This long spatial interaction allows for    strong absorption, which delivers a high signal to noise ratio, while simultaneously minimising the density of the molecules, which thereby minimises  the pressure broadening.    Furthermore, it is a common desire, in both spectroscopic studies and frequency standards, to desire access to the intrinsic spectroscopic properties of the molecule unhindered by    Doppler broadening and alignment~\cite{Argence} effects.   In this regard, the   fibre approach offers a great advantage over cell-based techniques because it automatically ensures a stable and effective overlap between the  counter-propagating pump and probe lasers beams required for saturated absorption techniques.  This high quality overlap arises because the vapour is stored inside an optical waveguide. The experiments reported below  demonstrate directly the quality of this overlap in the fibres. 
A final useful advantage of the iodine loaded HC-PCF is that the guiding and confinement within the small core of the fibre allows high intensities over long lengths for a small power inputs.  This allows deep saturation features to be achieved at small input powers which delivers a  substantial benefit for portable applications.

The most obvious  drawback of fibre-based vapour cells is the limited duration of the coherent light-molecule interaction because of the small diameter of the fibre core.   This limit causes an unavoidable line broadening (termed transit-time broadening) which   is of the order of   $\delta \omega_{TT} = v_{mp}/ \phi$ where $\phi$ is the HC-PCF core diameter and v$_{mp}$ the most probable velocity of the particles. The narrowest sub-Doppler bandwidths so far observed in HC-PCF have been in the range of 6-8\,MHz \cite{Hald, Knabe, chrispaper, PhilRB} and are limited by transit time broadening. It was necessary to choose either a large core (70\,$\mu$m) fibre \cite{Knabe}, apply anti-relaxation coatings \cite{PhilRB}, or make use of velocity selection effects through careful excitation strategies \cite{Hald,chrispaper} to obtain these narrow lines. 

This work uses a different approach by making use of a large mass molecule to reduce $v_{mp}$ for a given temperature gas.  The mass of the $^{127} I_{2}$ molecule is ten times that of acetylene $^{12}C_{2}H_{2}$ and 15 times larger  than ammonia $NH_{3}$ which gives a reduction in the potential transit time effects by a factor of 3-4 over those other vapours. This minimizes a key potential defect of fibre-based vapour cells.
%

In this paper we present iodine spectroscopy in the confined geometry of the hollow-core fibre and report, to our knowledge, the highest-Q spectral features yet observed in this environment.  We  compare these results to those from a conventional cell and illuminate some of the key differences.  Further, we will present a relatively simple theoretical model which is in good agreement with the observed features.  

\section{Experiment}

\begin{figure} [h!]
\centering\includegraphics[width=13cm]{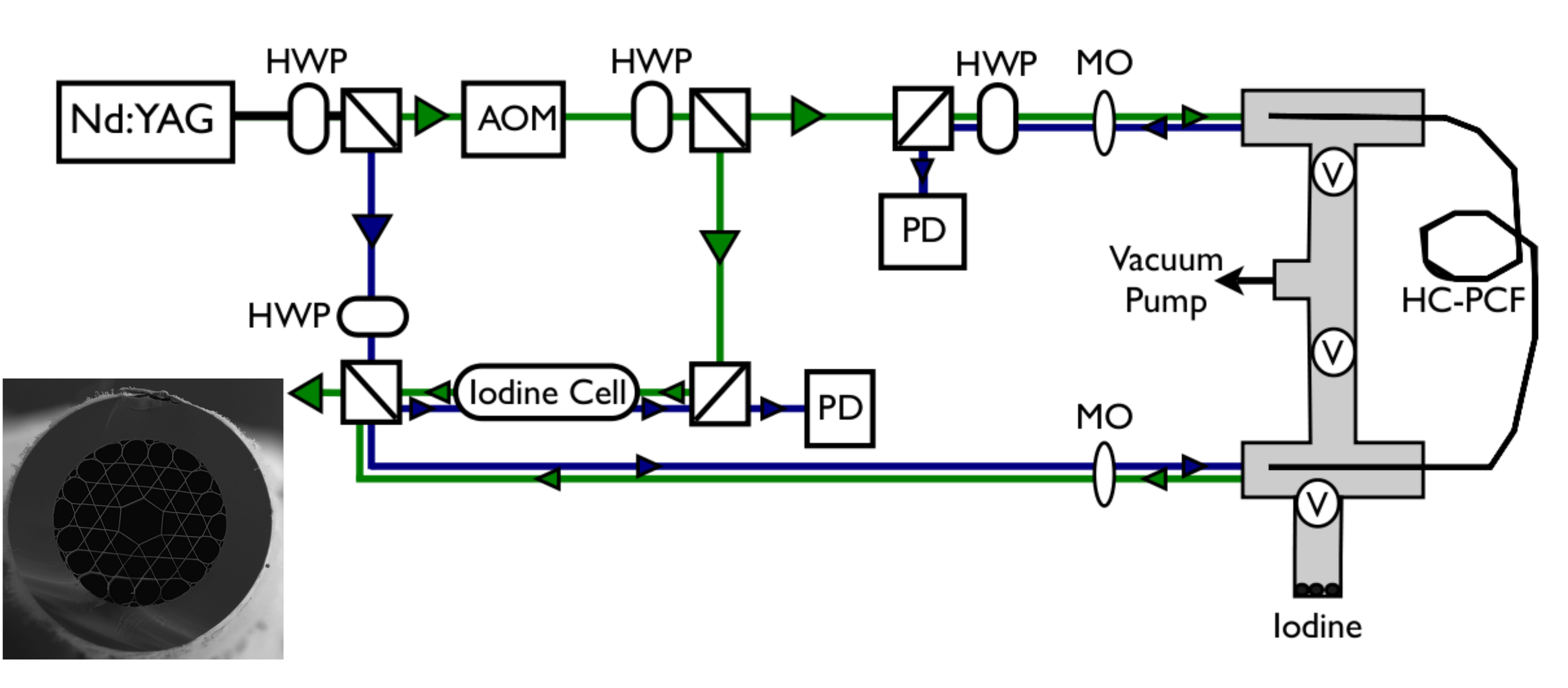} 
\caption{Optical setup with counter-propagating pump (green) and probe (blue) beams. MO - microscope objective, AOM - acoustic optical modulator, PD - photodiode, HWP - half-wave plate, V - vacuum valves, with the counter-propagating beams separated for ease of visualization. The inset shows a micrograph of the fibre used in this experiment.  The  diameter of the core is 24\,$\mu$m. } 
\label{optical}
\end{figure} 

Iodine presents a dense spectra of  ro-vibronic  transitions ranging   from green  through to the near-IR.  Each of these transitions consists of either 15 or 21 hyperfine lines due to interaction with the two nuclear spins ($I=\frac{5}{2}$ for each nucleus).  This hyperfine manifold is not properly resolved in the presence of Doppler broadening and thus in this experiment we have made use of  saturation spectroscopy to eliminate the inhomogeneous broadening.   In order to enhance the signal to noise ratio of the weak sub-Doppler spectral features we use a particular variant of modulation-transfer spectroscopy (MTS) in which we amplitude modulate the pump and synchronously detect the resulting amplitude modulation on the probe signal.  


A block diagram of the optical arrangement in shown in Fig.~\ref{optical}. The HC-PCF is mounted in a glass vacuum system with the majority of the 1.3\,m fibre outside the vacuum. A micrograph of the kagome-type~\cite{kagome} cross-section of the HC-PCF   is shown as an inset in Fig.~\ref{optical}. The fibre exhibits a loss of 0.7\,dB/m at our operating wavelength. A Viton cone is used to produce a compressive seal between the fibre and glass at the two points where the fibre enters the vacuum system. The open end of the fibre is suspended inside the vaccum a few millimeters behind an optical window so that there is  optical access to the core of the fibre during the filling phase.  Light is coupled into each end of the fibre using a $4\times$ microscope objective mounted on a three-axis translation stage. We obtain a total fibre transmission, including input and output losses together with fibre and window attenuation,  of around  60\%. 
Solid iodine  is kept in a separate section of the vacuum system at room temperature (equilibrium vapour pressure $\sim 36\pm1$\,Pa)  and is released into the HC-PCF after the vacuum system has been evacuated  down to a pressure of of $6\times10^{-6}$ Torr. The vacuum valves consist of a glass housing with a teflon valve body as this was found to minimize unwanted chemical interactions with the iodine vapour. 

 A  narrow-linewidth frequency-doubled  Nd:YAG laser provides 20\,mW of laser light  that is   tunable over $\sim$60\,GHz by controlling the crystal's temperature. This output light is split and counter-propagating pump and probe beams are sent through both the iodine-filled fibre and a  10\,cm long traditional iodine gas cell operated concurrently as a reference.  The transmitted probe light is separated from reflected pump light, using a polarizing beam-splitter and careful polarization tuning,  and then sent to a monitoring photodetector.  Despite this care, scattered pump light still caused interference on the detector - to overcome this the pump light was shifted by  200\,MHz using an Acousto-Optical Modulator (AOM). This same AOM  provided the amplitude modulation necessary for the MTS technique: it modulated the pump with 100\% depth at a 9.2\,kHz rate.  The output voltage of the photodiode was monitored by two instruments as the laser frequency was slowly swept: (i) the average voltage was measured to provide an estimate of the conventional linear absorption of iodine, (ii) the probe amplitude modulation transferred from  the pump modulation was demodulated with a lock-in amplifier to measure the magnitude of the sub-Doppler features.


\section{Theory}

\begin{figure} [h!]
\centering\includegraphics[width=5cm]{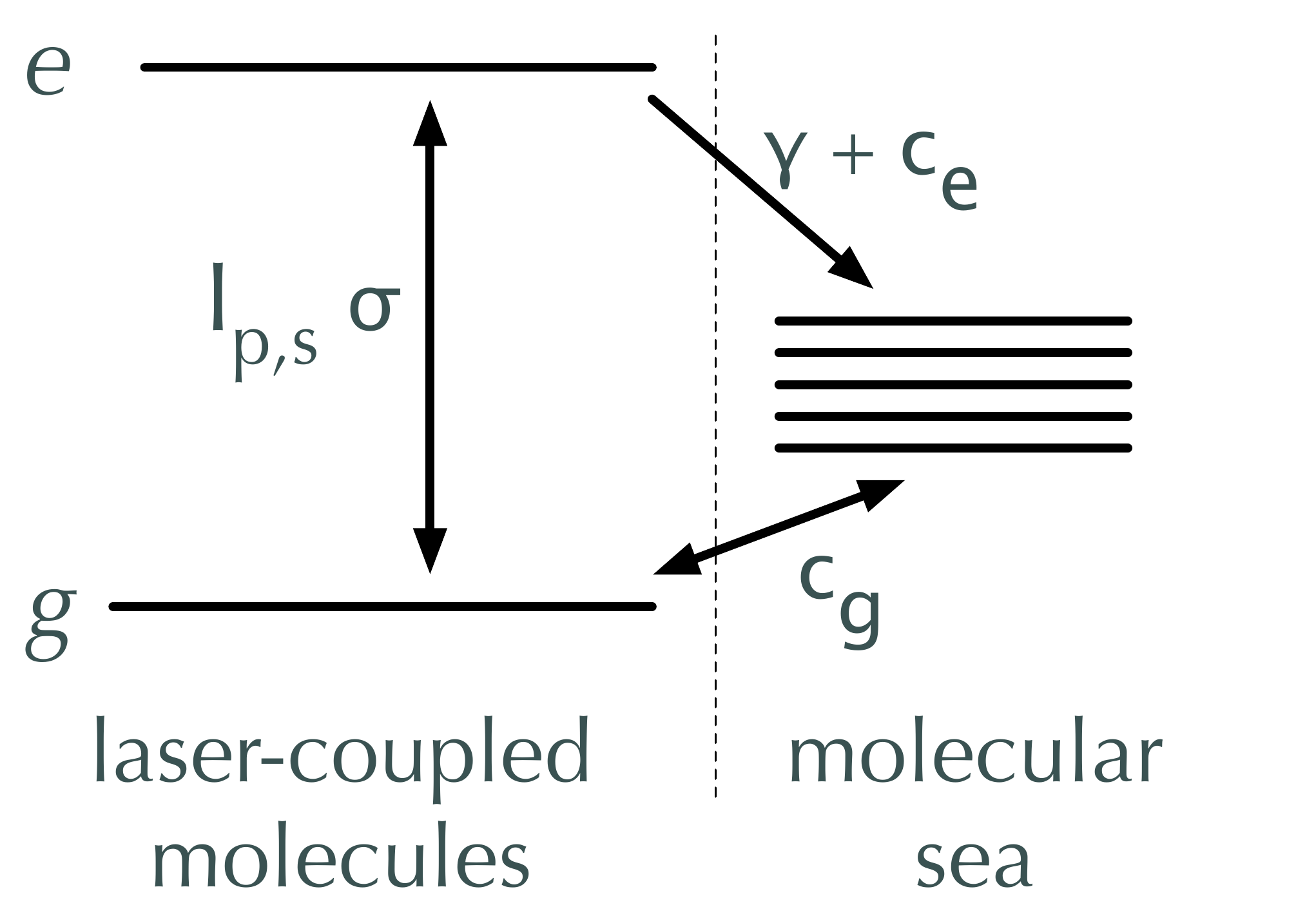} 
\caption{Schematic plot of the ``open''  two-level model of iodine molecules used in the theory presented here. Counter-propagating laser light of power $I_{p,s}$ drives stimulated transitions between the ground, $g$, and excited, $e$, states. Collisions with molecules in the ground and  excited states (at rates $c_{g}$ and $c_{e}$ respectively)  exchange  molecules between the laser-coupled and uncoupled states.    } 
\label{leveldiag}
\end{figure} 

We  use a simple ``open'' two level rate-equation model to describe the saturation spectroscopy of iodine (see Fig.~\ref{leveldiag}). This rate-equation approach will be valid  where the decoherence    is sufficiently   high   to keep the atomic coherences small (adiabatic approximation).  
We include in our rate equations collisional processes  that knock the molecules out of resonance with the laser light - either because their velocity has been changed or because their rotational, vibration or electronic state is modified. In either case the  molecules are considered to return  to a large ``sea'' of non-interacting iodine molecules~\cite{schenzle}. We have modelled these collision processes as a population loss from their  respective levels as population losses from their respective level  because we are performing  high-resolution spectroscopy in which even the weakest collisions modify the velocity or molecular state so that the molecule is out of resonance with the laser~\cite{banash,rose}. We note the close agreement between theory and experiment below which lends credence to this approach.

  We define rate equations for the excited ($e(t)$) and ground ($g(t)$) states in the reference frame of the molecule and imagine it to be interacting with two different colour laser signals representing a probe beam (labelled with ``s'' subscript) and a counter-propagating pump (labelled with ``p'' subscript):
\begin{equation}
\dot{e}(t) =   \frac{(g - e) I_p \sigma }{1 +  \frac{ {\delta_p}^2}{ (w_{0}/2)^{2}}  } +\frac{(g - e) I_s \sigma  }{1 +  \frac{ {\delta_s}^2}{ (w_{0}/2)^{2}} }  - e (c_e + \gamma) 
\label{estate}
\end{equation}

\begin{equation}
\dot{g}(t) =    (\rho \left[ \delta_{v} \right]-g) c_g - \frac{(g - e) I_p \sigma }{1 +  \frac{ {\delta_p}^2}{ (w_{0}/2)^{2}}  } - \frac{(g - e) I_s \sigma  }{1 +  \frac{ {\delta_s}^2}{ (w_{0}/2)^{2}} } 
\label{gstate}
\end{equation}

 where  $\delta_p =\delta_{a}- ( \delta_{AOM}+\delta_l + \delta_v)$ and $\delta_s = \delta_{a}- ( \delta_l - \delta_v)$ represent the pump and probe laser detuning frequencies from the molecular resonance frequency, $\delta_{a}$, translated into the frame of a molecule moving at a velocity $v = \delta v/k$.  $\delta_{AOM}$ takes account of the the fixed AOM frequency.  We include different collisional rates for the excited ($c_{e}$) and ground state ($c_{g}$) because  collisional rates have been shown to be highly state-dependent for iodine-iodine collisions~\cite{banash,schenzle,sleva}.  We include a population loss term for spontaneous emission from the excited state ($\gamma$) although due to the multitude of possible decay routes we have not included a corresponding gain in the ground state due to this process. On the other hand we have included a gain term ($\rho\left[\delta_{v} \right] c_{g}$) in the ground state representing  scattering from a ``sea'' of uncoupled states into the laser-coupled ground state.  The inhomogenous broadening is contained in the velocity detuning dependence of the $\rho\left[\delta_{v} \right] $ term. 
The pumping rates $I_{p,s}$ are given in terms of a photon rate per area while the scattering cross-section, $\sigma$, is defined such that $\sigma I_{p,s} $ is the  on-resonance photon scattering rate from a molecule irradiated with an intensity $I_{p,s}$ on the particular transition of interest. 
 All relaxation and pumping rates are expressed in units of s$^{-1}$, while  all frequencies are expressed in angular frequency units.
 %

The homogenous bandwidth of the molecule-light interaction, $w_{0}$, is set by the total coherent interaction time. Where the probe and pump beams are sufficiently large in diameter that transit-time broadening is not a consideration (e.g. in the conventional cell experiment reported here), the bandwidth will be $w_{0} = c_{e}+c_{g}+\gamma$ where we have ignored the effects of phase-interrupting collisions~\cite{banash}. However, when   transit-time effects are significant (as with the fibre experiment) then the  homogeneous bandwidth  is calculated using   the formalism   in Ref.~\cite{Borde}).  The early termination of the interaction between the probe light and molecules is modelled in the rate  equations above with an additional  effective collision that    adds equally to   both  $c_{e}$ and $c_{g}$ (since it terminates the light interaction whether we are in the ground or excited states).  In our experimental conditions,  the natural lifetime~\cite{Capelle} of the upper state ($1/\gamma \sim 1 \mu$s ) is much longer than the mean time between collisions for the excited state ($c_{e } > \gamma$) and so can usually be ignored.  

For comparison with the experiments below we wish to calculate the probe absorption.  The  definition of the probe absorption, $\alpha_{s}$ relates to the  change in intensity of the probe beam over an infinitesimal distance, $dz$ through a vapour, $dI_{s}/dz = \alpha_{s} I_{s}$.  Using energy conservation we can relate this intensity change   to the  energy scattered by the molecules in some volume of cross-sectional area, $A$, and length $dz$ as $ \left[ I(z+dz)-I(z) \right] A =  dz \, A \, e_{s} \times (c_{e}+\gamma) h \nu$   where $e_{s}$ is restricted to that fraction of the excited state population produced by the probe alone. To connect the rate of change of intensity to these concepts we utilize the fact that  $ \left[ I(z+dz)-I(z) \right] A \sim dI_{s}/dz A dz$. Finally, we  calculate $e_{s}$ by differencing  the excited state population in the presence and absence of the probe while the pump is held at a fixed power.



We  solve the rate equations in the steady-state  ($\dot{e}(t) =0, \dot{g}(t)=0$) to obtain relationships between the model parameters  and experimentally  observable features such as the absorption, saturation intensity and  the height and width of the Doppler-free features.  The intensity dependence of the  on-resonance ($\delta_{s}=0$) probe absorption can be calculated by setting the pump intensity to zero ($I_{p}=0$) and  integrating the steady-state solutions over the Doppler broadening implicitly contained in $\rho\left[\delta_{a} \right]$ to obtain:

 \begin{equation}
\alpha_{D}  =  \rho_{0} \sigma     \sqrt{\frac{1}{1+\frac{  I_{s} \sigma ({c_{e}}+\gamma +{c_{g}})}{{(c_{e}+\gamma)}{c_{g}}}}}
\label{satofDA1}
\end{equation}
where $\rho_{0} = \frac{\rho\left[\delta_{a} \right] \pi w_{0}}{2}$ is the population spectral density integrated over the homogeneous bandwidth. 
One notes that the probe absorption exhibits   the traditional   form:

 \begin{equation}
\alpha_{D}  =  \rho_{0} \sigma     \sqrt{\frac{1}{1+\frac{  I_{s} } {I_\mathrm{sat}}}}
\label{satofDA}
\end{equation}
if we interpret $I_\mathrm{sat}$ as = $ \frac{(c_e +\gamma)  c_g}{\sigma (c_e+\gamma + c_g)} \sim \frac{c_e  c_g}{\sigma (c_e+ c_g)} $ as the saturation rate for the linear absorption. Importantly for our analysis below we note that if there is a large disparity between the ground and excited state collisional rates then  the saturation rate will equal the slower of the two rates.  In contrast, the bandwidth, $w_{0}$, is equal to the sum of the collisional rates and hence will be approximately equal to the larger of the two collisional rates in that situation.  This different dependence of the bandwidth and saturation intensity allows us to discover a strong difference between the behaviour of the two collisional rates in the fibre and cell experiments. 


\begin{figure} [h!]
\centering\includegraphics[width=8cm]{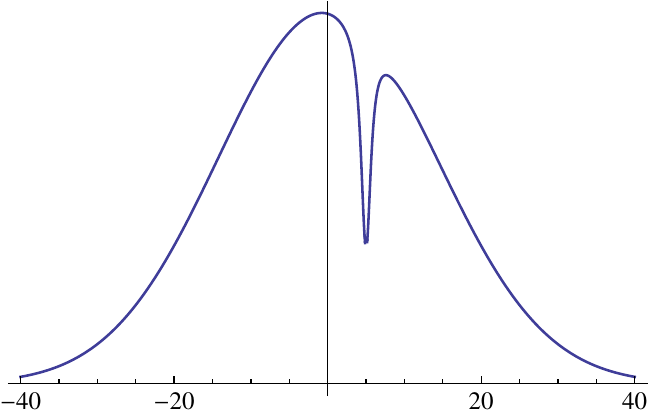} 
\caption{Probe absorption calculated in the presence of a saturating pump signal.    $I_{p}, \sigma, w_{0}, c_{g}, c_{e} $ have been arbitrarily set to 1, while  the inhomogeneous broadening is $\rho\left[\delta_{v} \right] = \exp(-\delta_{v}^{2}/\delta_{w})$ where $\delta_{v}$ is the offset frequency from the atomic frequency, $\delta_{a}$ and  the Doppler width, $\delta_{w}$, is set to be 20 times the homogenous bandwidth, $w_{0}$. The AOM frequency was arbitrarily set to 10 times $w_{0}$. The horizontal frequency axis is  normalized to $w_{0}$} 
\label{schem-doppfree}
\end{figure} 

In the presence of the pump one  sees a narrow feature in the probe absorption (see Fig.~\ref{schem-doppfree} for example).  This arises from the partial saturation of the ground and excited state population difference   by the   pump, which is   centred on $\delta_{p} = \delta_{s}$  (or equivalently $\delta_{l} = \delta_{a}-\delta_{AOM}/2$). For $I_{p},I_{s} < 100 I_\mathrm{sat}$ this feature closely follows  a Lorentzian shape: $\mathcal{L}(\delta) = A/(1+ 4 \delta^{2}/w^{2})$. The probe absorption observed far from this narrow feature is self-evidently equal to that calculated in Eq.~\ref{satofDA} since the pump and signals do not interact when tuned to different velocity classes.The reduced absorption at the centre of the saturation dip can be calculated to be:

 \begin{equation}
\alpha_{SD} = \rho_{0} \sigma \frac{1+ I_{p} / (2 I_\mathrm{sat})}{(1+ I_{p}/I_{\mathrm{sat}})^{3/2}}  
\label{dip}
\end{equation}
where we have used  the approximation that the probe itself is not causing a high degree of saturation of the iodine.  
  In this same limit, the effective contrast ($=1- \alpha_{SD}/\alpha_{D}$) of the  saturation absorption feature is:

 \begin{equation}
C   = 1- \frac{1+ I_{p} / (2 I_\mathrm{sat})}{(1+ I_{p}/I_{\mathrm{sat}})^{3/2}}  
\label{contrast}
\end{equation}
One   sees  that the contrast  will  reach 50\% for $I_{p}    \sim 1.1 I_{\mathrm{sat}}$. 
 The final key element of interest is the bandwidth of the saturation feature. We find that there is no simple closed-form expression for this although the expression below  is  within  1\% of the correct value over all realistic pump power values ($I_{p} < 100 I_\mathrm{sat}$): 

 \begin{equation}
w = w_{0} \sqrt{1+   I_{p}  / (\sqrt{2} I_{\mathrm{sat}})}
\label{FWHMequation}
\end{equation}
%
As expected the low power bandwidth   equals   $w_{0}$ while   additional  broadening is seen as the pump power exceeds   $I_{sat}$.
 
\section{Results}

\begin{figure}[tb]
\centering\includegraphics[width=13.5cm]{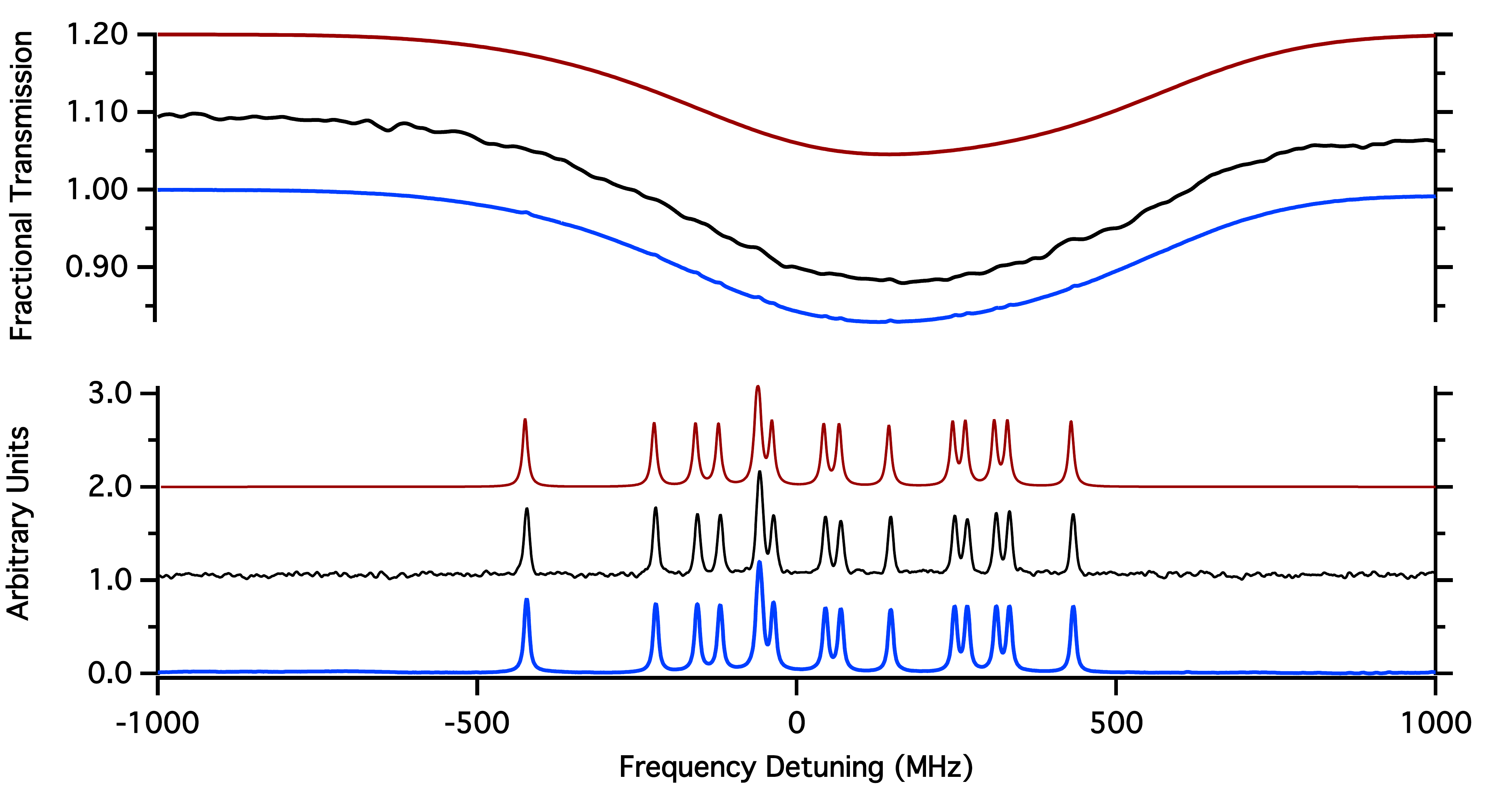} 
\caption{The P(142)37-0 Doppler-broadened (upper panel) and Doppler-free spectrum (lower panel)  recorded in the cell (lowest traces- blue), in the fibre (middle traces -black) and calculated using the theory presented in Sect. 3 (highest traces - red).  For the experiment the pump intensity was 41\,kWm$^{-2}$  and 25\,kWm$^{-2}$ for the fibre and cell respectively with probe intensities less than 20\% of that of the pump.  For the theory, we have used $w_{0} = 6$\,MHz; $\rho_{0} \sigma = 0.02$; $I_{p}/I_{sat} = 2$ and summed the response of all 15 hyperfine features using the known frequency spacings in Ref.~\cite{P142}. We have normalized the frequency axis to the centre of mass of the transition at   563.281\,GHz \cite{P142}.  The a1 hyperfine component is the lowest frequency component in the hyperfine spectrum near -424\,MHz. For clarity the lower traces have been offset by 1 unit, while the upper traces are offset by 0.1 units.}
\label{hfspect}
\end{figure}

Figure \ref{hfspect} shows the linear and sub-Doppler absorption spectra for the P(142) 37-0 iodine transition in the cell (10\,cm long) and fibre (1.3\,m) respectively using nearly identical pump intensities.  This is close to the maximum available intensity in the cell and near to the minimum used in   the fibre case (because of poor signal to noise at these low power levels).  The additional noise seen on the fibre data comes because the power is 600 times below that used in the cell along with some  vibrational noise which modulated the power coupled into the fibre core. 
From this figure it is evident that both measurements give extremely similar characteristics for the sub-Doppler features despite the confined geometry within the fibre core.  

As can be seen on Fig.~\ref{hfspect} both the fibre and cell have a maximum linear absorption of around 20\% on resonance.   The particular shape of the Doppler-broadened curve arises because it is the sum of 15 underlying hyperfine components each of which is broadened by the Doppler width of  437\,kHz.  In light of the small linear absorption we have made no allowance for axial variation in intensity in our model.    Our total end-to-end transmission loss through the fibre is around 60\% of which 20\% is associated with the separately measured fibre attenuation (1.3\,m, 0.7\,dB/m). The   rest of the loss is attributed to input coupling losses associated with mode-matching inefficiencies.  For all fibre intensities quoted below we   have upwardly corrected the quoted intensities by a factor of 1.5 with respect to the measured input intensity  to allow for the estimated input loss.

We recorded a large number of   spectra of the type shown on Fig.~\ref{hfspect} for both cell and fibre where we     varied  either the pump or probe intensity. We    used the known   frequency spacing between the a1 and a15 hyperfine components (855\,MHz \cite{P142}) to provide a  frequency axis for the measurements of the bandwidths of the individual hyperfine components. From these spectra  we derived the on-resonance absorption coefficient together with the contrast and bandwidth of the Doppler-free   features  as a function of     intensity.  Below we will examine each of these measurements and compare them to the theoretically expected result.

\begin{figure} [tb]
\centering\includegraphics[width=13cm]{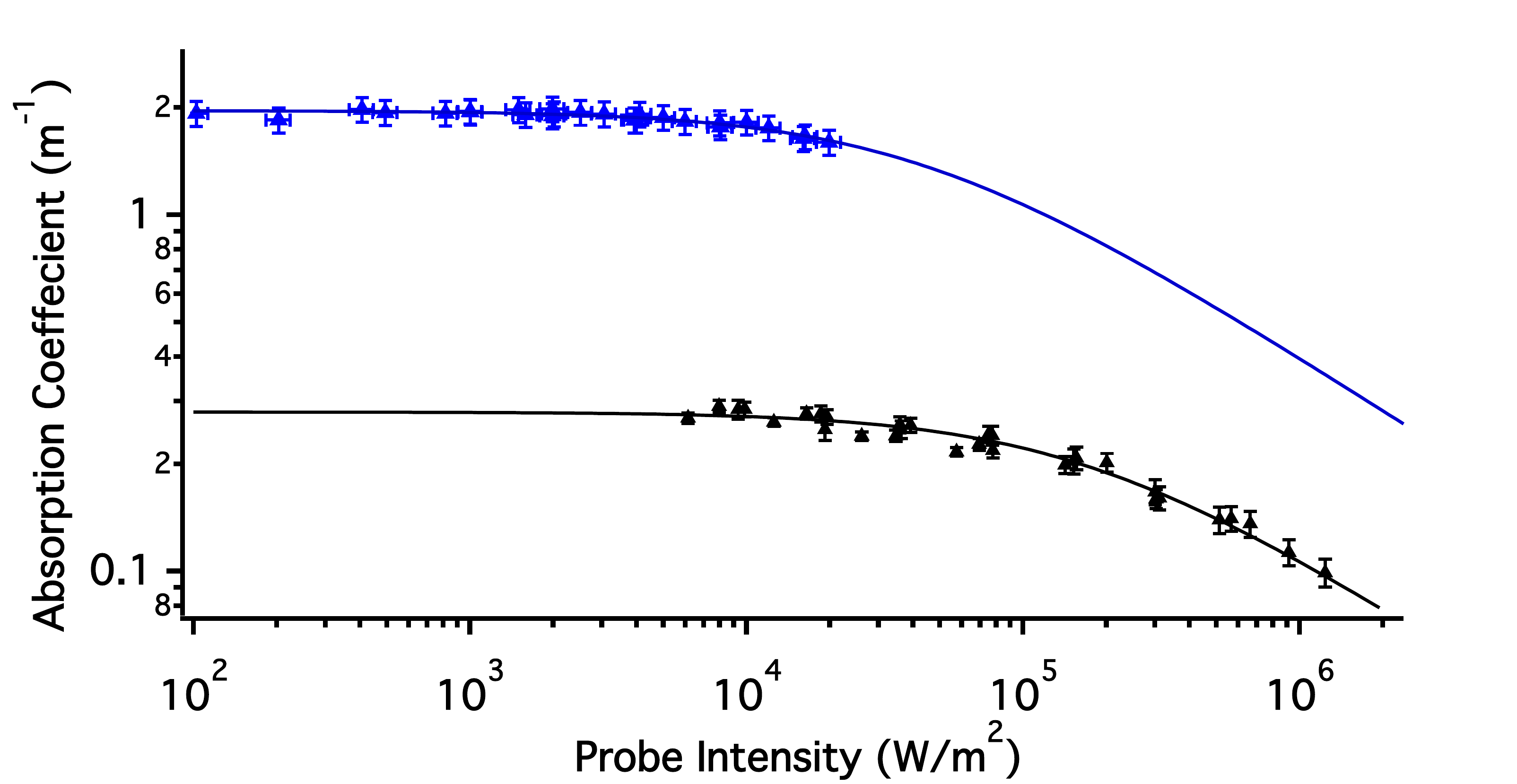} 
\caption{Linear Absorption of a1 component  of P(142) 37-0 line as a function of probe intensity. Black markers show measurements from HC-PCF while the blue markers show measurements from cell.  Solid curves in both cases are fits following the form of Eq.~\ref{satofDA}} 
\label{linabs}
\end{figure}

 Figure ~\ref{linabs}   presents the  absorption coefficient for the cell and fibre as a function of the probe intensity.   The   ratio of the low power absorption coefficients is $\sim 0.15$ and if we combine this with the known equilibrium vapour pressure of iodine at room temperature in the cell  then we can calculate the vapour pressure in this particular fibre load  at around $5.4 \pm 0.8$\,Pa. The solid lines shown on  Fig.~\ref{linabs} are a fit using Eq.~\ref{satofDA} and from this we extract a saturation intensity of $42 \pm 4$ kW/m$^{2}$ and $168 \pm 14$\,kW/m$^{2}$ for the cell and fibre respectively. This result indicates that effective collisional rate,  $\frac{c_e  c_g}{(c_e+ c_g)} $, differs substantially in spite of the similar decoherence processes  in the  fibre and cell.

\begin{figure}[tb]
\centering\includegraphics[width=13cm]{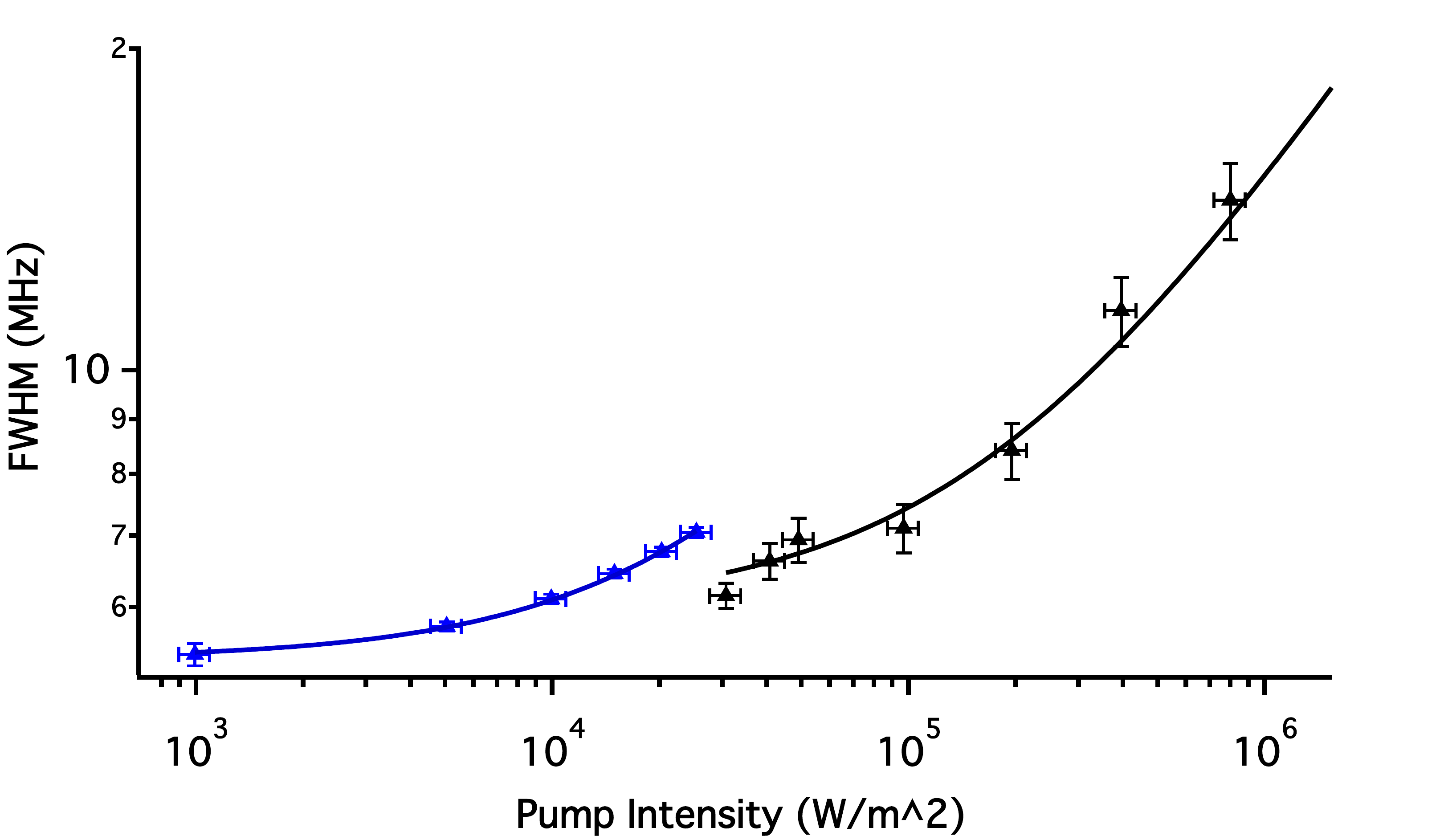} 
\caption{Bandwidth (FWHM) of the a1 hyperfine component of the P(142) 37-0 iodine transition measured using a 10cm long traditional glass iodine cell (blue markers), and in the iodine loaded HC-PCF (black markers), against total intensity. Solid curves in the two cases are fits to the data following the form of Eq.~\ref{FWHMequation}.} 
\label{FWHMfig}
\end{figure}

To further emphasise this point, we examine the bandwidth of the sub-Doppler features  seen on the lower panel of Fig.~\ref{hfspect} over a range of different pump powers.  In each case   the probe power was held  below 20\% of the pump power so that it  did not influence the bandwidth of the feature. We fit a   Lorentzian  function to the  a1 component in each trace and have summarised  these measurements on Fig~\ref{FWHMfig}. We also present on this figure solid curves of the form of  Eq.~\ref{FWHMequation}) and it can be seen that these are in close agreement with the experiment.  The fits yield nearly near-identical ``zero-intensity''  bandwidths of $5.4\pm0.02$\,MHz for the cell and $6\pm0.3$\,MHz in the fibre.  On the other hand,  the saturation intensities for the two cases are once again quite disparate  giving   saturation intensities of  $I_{sat}=24.0\pm0.5$\,kWm$^{-2}$  and $I_{sat}=129\pm16$\,kWm$^{-2}$ respectively. The saturation intensities derived from the bandwidth data are substantially below  those derived from the linear absorption data:    we believe that this principally arises  because the Doppler-broadened absorption includes contributions  from 15 underlying transitions with different frequency detunings from the probe laser which lowers the effective intensity. In addition, for the fibre measurement there are always small intensity calibration errors at the level of 20\% due to undeterminable input coupling efficiencies for the pump and probe.

The fact that the `zero-intensity''  bandwidths are almost identical indicates that the effective decoherence rates for the cell and fibre are  similar; however,   the saturation behaviour is seen to be very different ($\sim 5$ times higher in the fibre than the cell). The origin of this apparent paradox lies in the fact that  the ratio $c_{g}/c_{e}$ differs substantially  for the fibre and the cell. 

%

In the cell, the low-intensity bandwidth arises  from collisional broadening effects since transit-time broadening and lifetime broadening are  negligible in comparison.  The  dominant    broadening process in the cell is from iodine-iodine collisions (self-broadening).  While there  is a  degree of state-dependant variation (more than a factor of 2) in the measured broadening coefficients  we find that this measurement  is in reasonable agreement with previously measured  values~\cite{Phillips, Hiller,fang2006,Masiello}.  In the fibre the iodine pressure is only 15\% of that of the cell  (from the linear absorption coefficient measurement at low intensities), which leads to an expected   collisional self-broadening of just 0.8\,MHz, which is in contradiction with its measurement. However, in the fibre we have an additional broadening process arising from the transit of the molecules across the narrow optical mode.  The total bandwidth arising from these combined   broadening processes depends upon the regime of observation, which is categorised by the dimensionless parameters~\cite{Borde, Hald, Chardonnet}: $\eta = \Gamma r/v$ and $\theta =\Omega r/v =  \sqrt{I_{p}/(2 I_{sat})} \eta$, where $r$ is the probe beam of 1/e radius, $r$, and $v$ is  the most probable thermal velocity. $\eta$ can be thought to  represent   the number of collisions that occur in a typical beam crossing time, while $\theta$ is the number of Rabi cycles that occur in a typical beam crossing time.  For our ``zero-intensity'' bandwidth measurement $\theta \ll1$ and $\eta \sim 0.3$ ($\Gamma \sim 0.8\,MHz, r \sim 9 \mu$m, $v \sim  139$\,ms$^{-1}$.  The predicted bandwidth~\cite{Borde,Chardonnet,Hald} (full-width at half maximum) in this regime is  given by $w_{0} = 0.48 \eta^{0.5} v/r \sim 4.2$\,MHz, while the experimental measurement is just  $\sim  1.8$\,MHz larger than this. We believe that the origin of this small additional bandwidth, corresponding to just 0.8\,MHz of additional collisional broadening, comes from background gas atoms (water/nitrogen/oxygen) that are loaded into the fibre along with the iodine. Lending support to this conjecture is an observed slow increase in the measured bandwidth ($\sim 7\%$ per hour)  after the loading process. The unwanted gas load is associated with  the poor pumping efficiency enforced by the unfortunate aspect ratio of the optical fibre core together with the use of teflon and Viton in the rest of the vacuum system, which prevents an effective pre-baking of the vacuum system.  Assuming this background gas to be mainly air we estimate a background pressure of   approximately  17\,Pa \cite{FLETCHER,Comstock} at the instant this data was taken.  If it were possible to eliminate this background gas using an improved  vacuum apparatus, while also reducing the iodine pressure by a factor of 5, then it should be possible to obtain a  feature bandwidth    of just   $\sim 2.6$\,MHz  limited mostly by transit-time broadening with  a small  residual collisional component\cite{Borde}. A fibre of $\sim 5$\,m would yield similarly strong absorption features as those reported here. 

We note that the  termination of coherent light interaction by the transit-time effect is state-independent and thus   will add  equally to the effective collision rates for the excited and ground state.  In addition, collisions with background gas atoms are expected to show a similar collisional cross-section for both excited and ground state collisions~\cite{Brand}.  This combination suggests that for the fibre we might expect $c_{g} \sim c_{e}$. On the other hand,    iodine-iodine   collisions  produce strong quenching and perturbation of the  excited state molecules while leaving ground state molecules relatively unperturbed i.e $c_{g} < c_{e}$ for a iodine-iodine collision~\cite{banash,Hiller}.  We believe that these effects explain the strong difference observed in the  saturation intensity of the fibre and cell in spite of  the total decoherence rates   being  similar for the two cases.  If $c_{e,\mathrm{cell}}/c_{g,\mathrm{cell}} \sim 0.2 c_{e,\mathrm{fibre}}/c_{g,\mathrm{fibre}} $  then we would expect an $I_{sat,\mathrm{cell}} \sim 0.2 I_{sat,\mathrm{fibre}} $ in accord with our results. In different words, the  low rate of population decay from the molecular ground state in the cell (due to the low iodine-iodine collision rates for this state~\cite{banash}) leads to a bottleneck in the interchange of laser-coupled molecules with the background sea.  This leads to an easily saturated media in comparison to the fibre despite the fact that the total decoherence in the two cases is almost identical.


To provide further confirmation of  our understanding of the system we can  examine the effect of pump intensity on the depth of the P(142)37-0 a1 saturation feature.
%
Within a single Doppler-broadened transition (see Fig.~\ref{hfspect}) that  there are a  manifold of possible transitions (15 hyperfine levels in this case).  Each of these other transitions   contributes to the  Doppler-broadened absorption at the location of the a1 hyperfine component and thus influences  the apparent contrast of the feature.  In order to make a meaningful   comparison between  theory and experiment we calculated the influence of the other hyperfine components at the location of the a1 component and corrected the measured contrast for this factor.  A numerical calculation using the known spacings and relative strengths of the hyperfine components~\cite{P142} results in a correction to the measured contrast of a factor of 1.78.

%


\begin{figure}
\includegraphics[width=14cm]{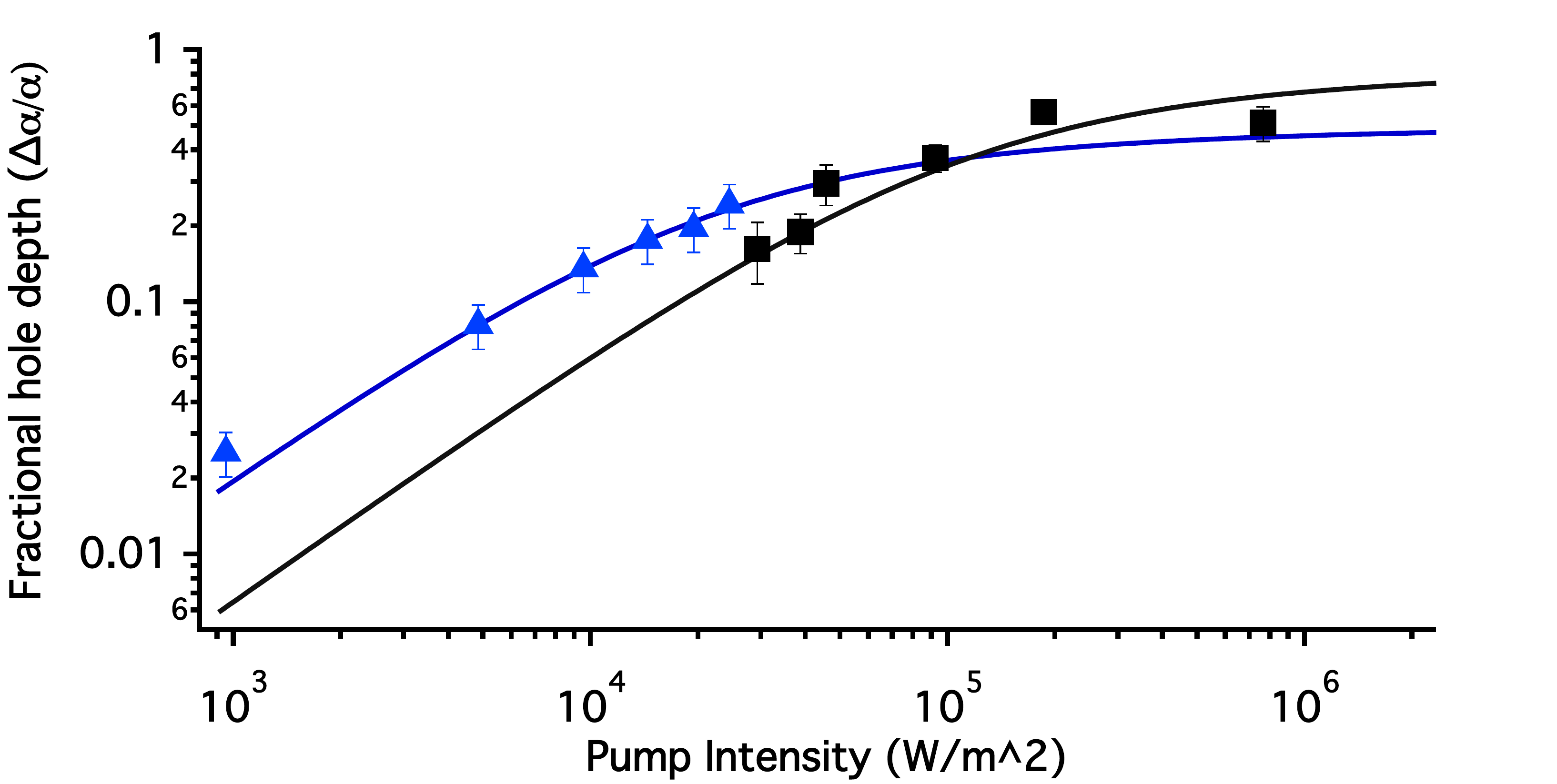}
\caption{P142(37-0) iodine a1 hyperfine transition hole depth versus pump intensity in the cell (low intensity data) and the HC-PCF (high intensity data). The probe was maintained at $\sim 20\%$ of the pump in both cases. The solid lines show fits of the form shown in Eq.~\ref{contrast}.}
\label{alphas}
\end{figure}
%


Figure \ref{alphas} shows the measured contrast with the corrective factor  applied. The solid lines on this figure are of the form shown in  Eq.~\ref{contrast}. We added an additional overall pre-factor to Eq.~\ref{contrast} to take into account  any imperfections in the beam overlap between pump and probe. This imperfection can arise from  attenuation of the pump and probe beams due to absorption through the vapour, or by the fibre, as well as to poor  spatial overlap of the two beams.  We have held the   saturation intensities fixed at those derived from the bandwidth data shown on Fig~\ref{FWHMfig}. 
 The pre-factor   was found to be 48\% for the cell data, while it is around 82\% for the fibre data. The excellent mode-matching of the pump and probe beams in the fibre leads to the factor of two improvement in   efficiency for   the fibre.

\section{Conclusion}
This paper has demonstrated   loading of a hollow-core fibre with $^{127} I_{2}$. We have presented Doppler-free and Doppler-broadened spectroscopic measurements on a particular iodine transition at 532\,nm ( P(142) 37-0 ). At room temperature, the a1 hyperfine component of this transition shows  a low power bandwidth of 6\,MHz in the HC-PCF which is  15\% larger than that observed in a traditional gas cell at the same temperature. The  bandwidth observed in this confined geometry compares favourably to the narrowest bandwidths measured for acetylene vapour in HC-PCF~\cite{Hald} and of Rubidium~\cite{chrispaper}. We have shown that the current bandwidth is limited by  collisions  with background gas atoms and we predict that by improving the vacuum and iodine loading system we could   achieve a bandwidths below 3\,MHz.  We have shown that the the fibre approach leads to an excellent efficiency for saturated absorption spectroscopy because of the high mode-overlap between the required pump and probe beams.  The Q-factor for this transition is $9 \times 10^{7}$, which we believe to be the highest observed in hollow-core fibre.

\section{Acknowledgements}
The UWA authors would like to thank the Australian Research Council for supporting this research through DP0877938 and FT0990301 research grants. We would like to thank Chris Perrella for insightful remarks on the manuscript and the rest of the group for creating a stimulating environment.  Finally, we would like to thank Jan Hall for his highly useful contributions in the early stages of this project.

\end{document}